\documentclass[superscriptaddress,10pt,aps,pra,twocolumn,longbibliography]{revtex4-2}
\usepackage{float}
\floatstyle{boxed}
\usepackage{array}
\usepackage{graphicx}
\usepackage{amsbsy}
\usepackage[utf8x]{inputenc}
\usepackage{epstopdf}
\usepackage{amsmath,amssymb,amsfonts,amsthm}
\usepackage{indentfirst}
\usepackage{soul}
\usepackage[T1]{fontenc}
\usepackage[dvipsnames]{xcolor}
\usepackage{url}
\usepackage[colorlinks]{hyperref}
\usepackage{array}
\usepackage{bbm}
\usepackage{dsfont}
\hypersetup{
    plainpages=true,
    breaklinks=true,
    hypertexnames=false,
    pageanchor=true,
    colorlinks=true,
    linkcolor={blue},
    citecolor={red},
    urlcolor={blue},
    anchorcolor={black}
}

\newcommand{\figref}[1]{\mbox{Fig.~\ref{#1}}}

\renewcommand{\eqref}[1]{\mbox{Eq.~(\ref{#1})}}

\newcommand{\be}{\begin{equation}}
\newcommand{\ee}{\end{equation}}
\newcommand{\bea}{\begin{eqnarray}}
\newcommand{\eea}{\end{eqnarray}}

\begin{document}
\title{Covariant formulation of the Berry connection in non-Hermitian systems}
\author{Ievgen I. Arkhipov}
\email{ievgen.arkhipov@upol.cz}
\affiliation{Joint Laboratory of
Optics of Palack\'y University and Institute of Physics of CAS,
Faculty of Science, Palack\'y University, 17. listopadu 12, 779 00
Olomouc, Czech Republic}

\begin{abstract}

Non-Hermitian systems exhibit spectral and topological phenomena absent in Hermitian physics; however, their geometric characterization remains subtle due to the intrinsic ambiguity of biorthogonal eigenspaces. Since left and right eigenvectors are not related by Hermitian conjugation, the associated Berry connection is generally nonunique, leading to complex geometric phases and ambiguously defined holonomies.
Here we formulate a covariant geometric framework for non-Hermitian quantum systems based on the metric structure of the underlying Hilbert space. We show that, in the quantum regime with continuous state evolution, the conventional Berry connection and the associated Berry holonomy over closed parameter-space loops can be consistently defined only in the pseudo-Hermitian limit, where the spectrum is real. For generic non-Hermitian Hamiltonians with complex spectra, the relevant geometric object is instead the Aharonov--Anandan holonomy associated with cyclic evolution in projective Hilbert space.
Within the pseudo-Hermitian regime, we construct a unique Hermitian Berry connection that is covariant under arbitrary ${\rm GL}(N,\mathbb C)$ frame transformations and reduces to the standard Berry connection in the Hermitian limit. The resulting formalism separates the intrinsic geometry of the Hamiltonian eigenspace from contributions arising from the parameter dependence of the Hilbert-space metric, revealing that the conventional biorthogonal formulation generally mixes these distinct geometric effects. Consequently, geometric phases, synthetic gauge fields, and topological characteristics commonly attributed to non-Hermitian eigenspace geometry may, in part, originate from the underlying metric structure. Our framework therefore provides a consistent geometric foundation for Berry phases, non-Abelian holonomies, and topological invariants in non-Hermitian quantum systems.

\end{abstract}

\date{\today}

\maketitle

%----------PRELIMINARIES
\section{Introduction}

Berry connection, geometric phase, holonomy, and the topology encoded in Bloch bands play a central role in modern quantum physics, underpinning phenomena ranging from charge pumping and the quantum Hall effect to topological insulators and superconductors~\cite{Berry1984_BP,Simon1983,Thouless1982,Xiao2010,Hasan2010,Qi2011,Ozawa2019}. Their extension to non-Hermitian systems has recently attracted considerable attention owing to the emergence of spectral and topological structures absent in Hermitian physics~\cite{Ashida2020,Gong2018,Kawabata2019,Bergholtz2021}.

Non-Hermitian Hamiltonians arise in a wide variety of physical platforms, including photonic~\cite{Guo2009,Peng2014,Doppler_2016,Hasan2017,Roy2020,Soleymani2022,Ergoktas2022,Qu2026,Wang2026},  mechanical~\cite{Bender_2013}, optomechanical~\cite{Xu2016,Jing2014} and electrical~\cite{Schindler2011,Ghatak2019,Schomerus2020,Helbig2020} systems, acoustic and phononic lattices~\cite{Zhu2014,Fleury2015,Ding2016,Shi2016,Huang2024}, cold-atom systems~\cite{Takasu2020,Ding2021}, and circuit-QED architectures~\cite{Naghiloo19,Parto2020} to name a few. These systems exhibit phenomena such as exceptional-point topology, the non-Hermitian skin effect, and non-Bloch band structures~\cite{Lee2016,Yao2018,Kunst2018,Borgnia2020,Okuma2020,Gong2018,Kawabata2023}, thereby revealing geometric and topological features with no direct Hermitian counterpart. See also extensive reviews on these and related topics~\cite{Ozdemir2019,El-Ganainy2018,Ozawa2019,Ashida2020,Bergholtz2021}.

A defining property of non-Hermitian Hamiltonians is the biorthogonality of their eigenspaces: right and left eigenvectors are generally independent and are not related by Hermitian conjugation~\cite{MOSTAFAZADEH_2010,Brody2013}. Consequently, eigenstates possess an intrinsic ${\rm GL}(N,\mathbb C)$ gauge freedom associated with independent local rescalings of left and right eigenvectors. While this freedom leaves spectral quantities invariant, it directly affects geometric objects constructed from eigenvector derivatives. In particular, the conventional non-Hermitian Berry connection admits four inequivalent biorthogonal forms depending on how left and right eigenvectors are paired~\cite{Garrison_1988,Shen2018,Singhal2023,Yang2024}. Since these connections generally transform differently under non-unitary gauge transformations, the resulting geometric phases and Berry curvatures become gauge dependent and, in general, complex valued.

Although such complex geometric phases are natural in classical and semiclassical dissipative systems, where the evolving field may physically amplify or attenuate along a trajectory~\cite{Silberstein2020,Montag2026,Singhal2023,Lane2025}, the situation is fundamentally different in the quantum regime.
When the wave function represents probability amplitudes, continuous and deterministic evolution requires preservation of the total state norm~\cite{Dirac1927,Barandes2025}, while dissipation and probability loss are associated with quantum jumps or conditional measurement processes~\cite{BreuerBookOpen,Minganti2020,Minganti2022}.
From this perspective, the common practice of enforcing normalization through {\it ad hoc} state renormalization~\cite{Silberstein2020,Singhal2023,Ozawa2025b,Montag2026,Behrends2025} may obscure the distinction between genuine eigenspace geometry and contributions originating from the underlying Hilbert-space metric.

These issues have motivated recent attempts to formulate Hermitian Berry connections for non-Hermitian systems~\cite{Yang2024,Yang2025,Zhang2019top_obst}. However, existing approaches remain restricted either to degenerate eigenspaces or to specific forms of the metric structure, and a general covariant geometric framework is still lacking.

In this work, we construct a unique Hermitian and covariant Berry connection for non-Hermitian quantum systems based on the metric structure of the associated Hilbert space. Most importantly, we show that, in the regime of continuous norm-preserving quantum evolution, the conventional Berry connection and the associated Berry holonomy can be consistently defined only for pseudo-Hermitian Hamiltonians characterized by real spectra. In contrast, for generic non-Hermitian Hamiltonians with complex spectra, the fundamental geometric object is no longer a Berry holonomy over closed parameter-space loops, but rather an Aharonov--Anandan holonomy associated with cyclic evolution in projective Hilbert space.

The resulting covariant formalism disentangles the intrinsic geometry of the physical eigenspace from metric-induced contributions associated with the parameter dependence of the Hilbert-space structure. As a consequence, geometric phases and synthetic gauge fields obtained within the conventional biorthogonal formalism may partially reflect metric effects rather than intrinsic properties of the Hamiltonian eigenbundle itself. The proposed framework therefore provides a consistent geometric foundation for Berry phases, non-Abelian holonomies, and topological invariants in quantum systems described by non-Hermitian Hamiltonians.

\section{Metric tensor and covariant derivative in the Hilbert space of non-Hermitian Hamiltonians}
Consider dynamics of a state $|\psi\rangle$ evolving under a non-Hermitian Hamiltonian (NHH) $H(\boldsymbol{\lambda})$ defined over the parameter space $\boldsymbol{\lambda}$ through the Schr\"odinger  equation (by setting $\hbar=1)$:
\begin{equation}\label{SE}
     i\partial_t |\psi\rangle = H(\boldsymbol{\lambda}) |\psi\rangle.
\end{equation}
To treat 
 $|\psi\rangle$
as a quantum state describing the probability amplitudes, we require its norm to remain unity. For a NHH this necessitates a positive-definite metric operator $\eta$ on its Hilbert space such that 
\begin{equation}\label{eta}
    \langle \psi | \eta | \psi \rangle=1.
\end{equation}
To ensure the latter the metric itself must obey the equation 
\begin{equation}
    \partial_t \eta = i\eta H - iH^\dagger \eta.
\end{equation}
Evidently, for Hermitian systems $\eta= I$. 
Because $\eta$ is Hermitian and positive-definite operator, it can be alternatively represented as 
\begin{equation}\label{S}
    \eta=S^\dagger S,
\end{equation}
for some invertible operator $S$, which is in general time-dependent. The invertibility of $S$ ensures that this decomposition is valid throughout the parameter space, except at spectral singularities, i.e., exceptional points (EPs), where the metric may diverge~\cite{MOSTAFAZADEH_2010}. Though  at the very EPs, the metric can still be defined using generalized eigenvectors~\cite{Ju2019}, or even in more general settings where the metric can be made everywhere smooth by appropriately modifying the Hamiltonian parameter space~\cite{Fring2022}. In this work, we restrict our analysis to regions where the metric is well-defined, as geometric observables, such as Berry phases and holonomies, require a smooth manifold structure away from spectral singularities. Accordingly, unless stated otherwise, throughout the text we implicitly assume that the map $S$ is smooth in these regions of interest. 

The operator $S$ plays the role of the vielbein within the framework of the Einstein elevator~\cite{Ju2022}, or the Dyson map~\cite{Dyson1956}. Because of this, one can effectively map the right vectors $|\psi\rangle$ of NHH to a vector 
\begin{equation}\label{psi-psi}
    |\phi^H\rangle\equiv S|\psi\rangle,
\end{equation}
corresponding to, in  general, a time-dependent Hermitian Hamiltonian~\cite{Fring2016}:
\begin{equation}\label{HH1}
    H^H=SHS^{-1}+i(\partial_tS)S^{-1}.
\end{equation} 
Note that the Hermitizing map $S$ possesses a unitary gauge freedom and is therefore not unique~\cite{Ju2022}. Indeed, under the transformation $S \to US$ with a unitary gauge $U$, the metric $\eta$ remains invariant. This unitary gauge freedom does not affect the conclusions of this work, since the resulting geometric phases and holonomies remain invariant under such gauge transformations.

One can choose a minimal $S$, whose parameter dependence stems solely from the metric, i.e., 
\begin{equation}
    S=\sqrt{D}W^\dagger, \quad \eta=WDW^\dagger,
\end{equation}
with $W\in U(N)$, and $D$ is a positive-definite diagonal matrix. 
We note that an alternative choice frequently encountered in the literature is $S=\sqrt{\eta}$, as adopted, for example, in Ref.~\cite{Zhang2019top_obst}. However, such a construction generally does not properly account for the parameter dependence of the metric eigenvectors themselves~\cite{Mostafazadah2018}, and may therefore obscure geometric contributions associated with rotations of the metric frame in parameter space~\cite{arkhipov2026a}.

In contrast to the conventional biorthogonal formalism, the vielbein $S$ enables a consistent definition of the inner product between states associated with different points on the extended parameter manifold
\begin{equation}\label{M}
{\cal M}
=
{\cal M}_\lambda
\times
\mathbb R_t,
\end{equation}
with coordinates
\begin{equation}\label{xmu}
x^\mu=(\boldsymbol{\lambda},t).
\end{equation}
For two points $x_1,x_{2}\in{\cal M}$, the physically meaningful overlap is defined as
\begin{equation}\label{x1x2}
\langle
\psi(x_1)
|
\eta(x_1,x_2)
|
\psi(x_2)
\rangle,
\end{equation}
where the corresponding two-point metric operator is
\begin{equation}
\eta(x_1,x_2)
=
S^\dagger(x_1)S(x_2).
\end{equation}

This contrasts with the conventional biorthogonal overlap
\begin{equation}
\langle
\psi^L(x_1)
|
\psi^R(x_2)
\rangle,
\end{equation}
where
\begin{equation}
\langle
\psi^L(x)
|
=
\langle
\psi^R(x)
|
\eta(x).
\end{equation}
Since, in general,
\begin{equation}
\eta(x_1)
\neq
\eta(x_1,x_2),
\end{equation}
the conventional biorthogonal construction does not provide a consistent inner product between states defined at different points of the extended manifold.

\begin{figure}[t!]
    \includegraphics[width=0.485\textwidth]{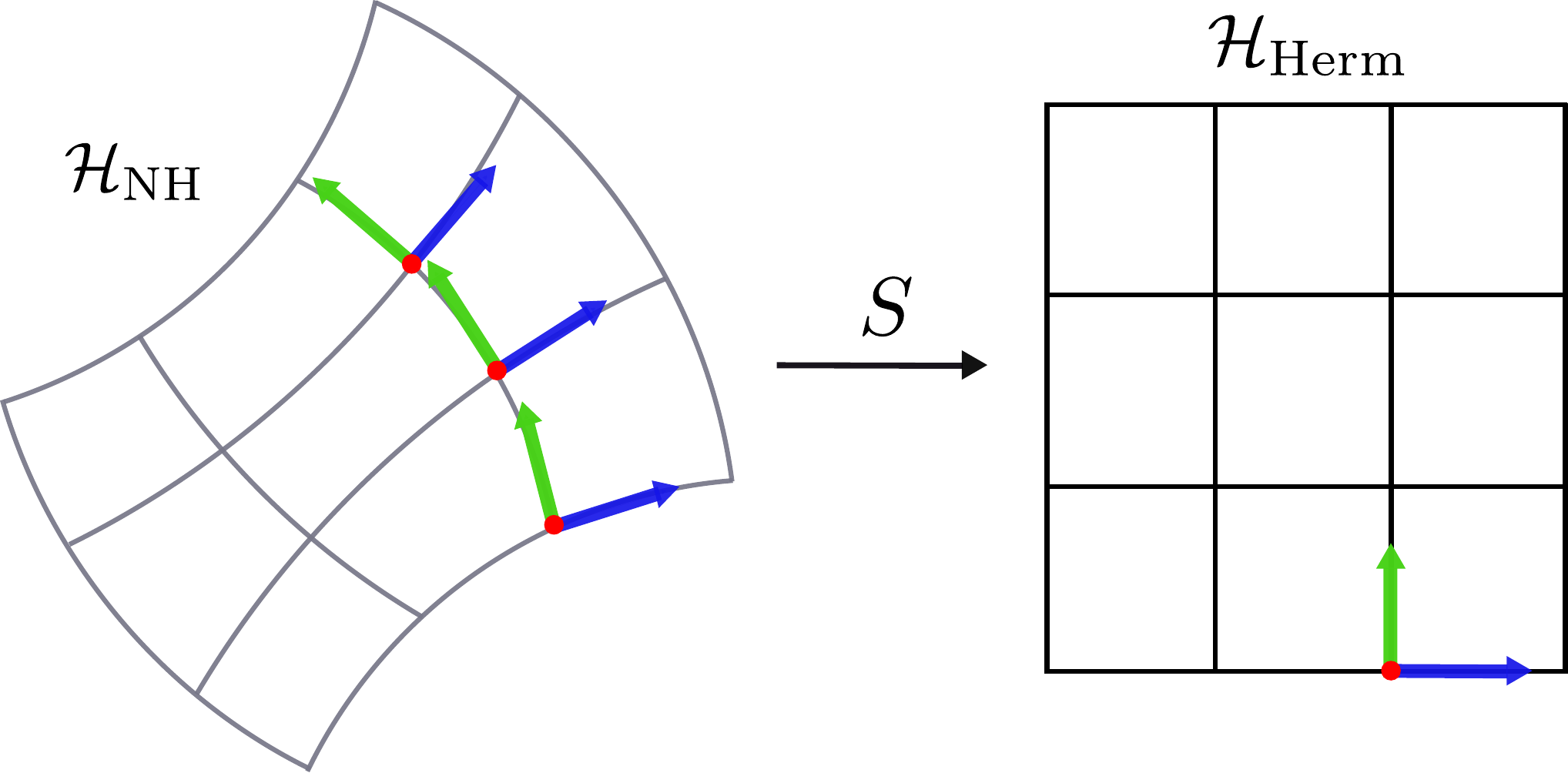}
    \caption{
    Symbolic illustration of the vielbein map $S$, according to \eqref{psi-psi}, which relates the deformed (e.g., stretched) Hilbert space ${\cal H}_{\rm NH}$ of the non-Hermitian Hamiltonian, equipped with the metric $\eta=S^\dagger S$, and characterized by nonequivalent local frames at different points, to the flat Hilbert space ${\cal H}_{\rm Herm}$ associated with the corresponding Hermitian Hamiltonian, which admits a single global frame.
    }
    \label{fig1}
\end{figure}

Consequently, in an $\eta$-deformed Hilbert space the inner product would vary across the system parameter space, rendering ordinary derivatives insufficient for comparing state vectors at different points. A covariant derivative is thus required~\cite{NakaharaBook}.
The construction of the covariant derivative is greatly simplified by exploiting the relation between the states of non-Hermitian and Hermitian Hamiltonians in \eqref{psi-psi}~\cite{Mostafazadah2018,arkhipov2026a}: 
\begin{equation}\label{D}
    D_{\mu}=\partial_\mu+S^{-1}\partial_\mu S,
\end{equation}
where the term
\begin{equation}\label{Gamma_def}
    S^{-1}\partial_\mu S=\Gamma^\mu,
\end{equation}
serves as the metric-compatible connection component on the $\eta$-deformed Hilbert space, and which obeys the equation: $\partial_\mu\eta=\eta\Gamma^\mu+\Gamma^{\mu\dagger}\eta$.
Furthermore, this connection is {\it curvature-free}, as it is generated by a pure-gauge structure (i.e., the Maurer-Cartan form) $S^{-1}d S$, implying $[D_{\mu'},D_\mu]=0$~\cite{NakaharaBook,Ju2024}.
Meaning that the Hilbert space, endowed with the metric 
$\eta$, is locally flat despite the seeming deformation (see \figref{fig1}).

\section{Aharonov--Anandan holonomy and Berry connection in pseudo-Hermitian limit}

\subsection{Aharanov-Anandan holonomy in non-Hermitian systems}
Owing to the existence of the Hermitizing map $S$ in~\eqref{psi-psi}, which establishes a one-to-one correspondence between non-Hermitian and Hermitian representations, the geometric aspects of non-Hermitian quantum evolution can be formulated consistently in the associated Hermitian frame. This provides a natural route for extending geometric notions, such as holonomy, to non-Hermitian systems in a manifestly covariant manner: the relevant geometric objects are first identified in the Hermitian representation and subsequently mapped back to the non-Hermitian frame.

The relation between a non-Hermitian Hamiltonian $H$ and its Hermitian counterpart $H^H$ is conveniently expressed through the metric-compatible (Dyson) temporal connection
\begin{equation}
\Gamma^t=S^{-1}\partial_tS,
\end{equation}
such that
\begin{equation}\label{HH}
H^H(t)
=
S\bigl(H+i\Gamma^t\bigr)S^{-1}.
\end{equation}
The operator inside the parentheses is pseudo-Hermitian and is related to the Hermitian Hamiltonian $H^H$ through a similarity transformation. In this sense, the Dyson connection $\Gamma^t$ compensates the non-unitarity associated with the explicit-time dependence of the metric stemming from the imaginary spectrum of the NHH.

Importantly, the Hamiltonian $H^H$ in~\eqref{HH} generally possesses explicit time dependence in addition to any dependence induced by the parameters $\boldsymbol{\lambda}(t)$. As a consequence, the instantaneous eigenspaces of $H^H$ do not, in general, define a vector bundle over the parameter manifold ${\cal M}_\lambda$ alone. In particular, even when the parameters $\boldsymbol{\lambda}(t)$ trace a closed loop, the evolution is not, in general, governed by a Berry holonomy associated with a parameter-space eigenbundle, which is the underlying assumption of the conventional Berry construction.

More specifically, in this setting time acts as an independent parameter, so that the quantum evolution is determined by the full trajectory of the state $|\phi^H(t)\rangle$ in projective Hilbert space rather than by a geometry induced solely from the parameter manifold ${\cal M}_\lambda$. Moreover, the explicit time dependence generated by the Hermitizing transformation need not correspond to a cyclic adiabatic driving protocol. Therefore, unlike, e.g., in Thouless pumping and related mechanisms, there is no generic justification for treating time as an additional periodic parameter and defining a Berry holonomy over closed trajectories in the extended `spacetime' manifold ${\cal M}$ in~\eqref{M}. Instead, the geometric characterization must be formulated directly in terms of the evolution of quantum rays in projective Hilbert space, naturally leading to the Aharonov--Anandan (AA) framework.

In other words, the appropriate geometric object is therefore the holonomy associated with cyclic evolution of the quantum ray. This leads naturally to the AA framework~\cite{Aharonov1987}, which assigns a geometric phase directly to closed curves in projective Hilbert space and does not rely on the existence of a parameter-space eigenbundle or on an adiabatic projection onto it.

To establish the corresponding AA holonomy, consider an arbitrary solution of the Schr\"odinger equation in the Hermitian frame
\begin{equation}\label{SEH}
i\partial_t|\phi^H(t)\rangle
=
H^H(t)|\phi^H(t)\rangle.
\end{equation}
Now, suppose that after a finite evolution time $T$ the state satisfies
\begin{equation}\label{gamma0}
|\phi^H(T)\rangle
=
e^{i\gamma}
|\phi^H(0)\rangle,
\end{equation}
so that the evolution defines a closed curve in the projective Hilbert space ${\cal P}({\cal H})$. The latter is obtained from the Hilbert space ${\cal H}$ through the projection map
\begin{equation}
\Pi:{\cal H}\rightarrow{\cal P}({\cal H}),
\end{equation}
which associates to every nonzero state vector its equivalence class
\begin{equation}
\Pi(|\phi^H\rangle)
=
[\,|\phi'^H\rangle\,]
=
\left\{
c|\phi^H\rangle:
c\in\mathbb C\setminus\{0\}
\right\}.
\end{equation}
The evolving state $|\phi^H(t)\rangle$ therefore defines a curve in Hilbert space,
\begin{equation}\label{CH}
{\cal C}_H:[0,T]\rightarrow{\cal H},
\end{equation}
whose projection
\begin{equation}
\Pi\circ{\cal C}_H:[0,T]\rightarrow{\cal P}({\cal H})
\end{equation}
defines the corresponding loop trajectory in projective Hilbert space. Consequently, for any given ${\cal C}_H$, one can define a Hamiltonian $H(t)$ that satisfies \eqref{SEH} for the normalized $|\phi^H\rangle$.

Following Ref.~\cite{Aharonov1987}, such cyclic evolution along ${\cal C}_H$ gives rise to the AA geometric phase,
defined as
\begin{equation}\label{gammaAA}
    \gamma_{\rm AA}=i\int\limits_0^T\langle\tilde \phi^H|d|\tilde\phi^H\rangle,
\end{equation}
where 
\begin{equation}
    |\tilde\phi^H(t)\rangle=\exp(-if(t))|\phi^H(t)\rangle,
\end{equation}
with the condition $f(T)-f(0)=\gamma$.
The dynamical phase is determined by the integral
\begin{equation}
\gamma_{\rm dyn}
=
-\int_0^T
\langle\phi^H(t)|
H^H(t)
|\phi^H(t)\rangle
\,dt
\end{equation}
such that the total phase $\gamma$ in~\eqref{gamma0} reads
\begin{equation}
    \gamma=\gamma_{\rm AA}+\gamma_{\rm dyn}.
\end{equation}

Transforming back to the non-Hermitian frame yields
\begin{equation}\label{Revol_AA}
|\psi^R(T)\rangle
=
S_T^{-1}S_0
\,e^{\displaystyle i\gamma_{\rm dyn}}
\,e^{\displaystyle i\gamma_{\rm AA}}
|\psi^R(0)\rangle.
\end{equation}
The prefactor $S_T^{-1}S_0$ accounts for the metric-induced frame transformation and guarantees consistency of the $\eta$-inner product according to~\eqref{x1x2}. In particular,
\begin{equation}
\langle\psi^R(0)|\eta(0,T)|\psi^R(T)\rangle \in {\rm U}(1),
\end{equation}
ensuring that the evolution is cyclic in projective Hilbert space equipped with the metric $\eta$.
In this sense, the closed curve ${\cal C}_H$ in the Hermitian representation induces a closed ray trajectory ${\cal C}_{NH}$ in the non-Hermitian frame in the $\eta$-deformed projective Hilbert space, and therefore defines a well-posed Aharonov--Anandan holonomy for the state $|\psi^R(t)\rangle$.

Now, expressing the dynamical and geometric phases via states in the non-Hermitian frame one attains
\begin{eqnarray}\label{gammaAAnh}
    \gamma_{\rm AA}&=&i\int\limits_0^T\langle\tilde\psi^R|D|\tilde\psi^R\rangle, \quad |\tilde\psi^R(t)\rangle=e^{-if(t)}|\psi^R\rangle,\nonumber \\
    \gamma_{\rm dyn}&=&-\int\limits_0^T\langle\psi^R|\eta(H+i\Gamma^t)|\psi^R\rangle dt, 
\end{eqnarray}
where the total covariant derivative is defined as $D = D_\mu dx^\mu$, in accordance with Eqs.~(\ref{xmu}) and (\ref{D}). Both phases are real in~\eqref{gammaAAnh}, consistent with their Hermitian-frame counterparts in~\eqref{gammaAA}.

Equations~(\ref{Revol_AA}) and~(\ref{gammaAAnh}) therefore define the AA holonomy associated with cyclic evolution generated by a generic NHH, irrespective of whether its spectrum is real or complex.
Importantly, the emergence of the AA holonomy is a direct consequence of requiring a covariant geometric description of the evolution. Since covariance must be maintained under the Hermitian--non-Hermitian correspondence generated by the Hermitizing map, the geometric phase is naturally associated with the evolution of quantum rays in projective Hilbert space rather than with a parameter-space eigenbundle alone, i.e., expressed via the Berry holonomy. Consequently, the appropriate geometric object for generic non-Hermitian systems is the AA holonomy. As shown below, the conventional Berry holonomy, in its covariant form, is recovered only in the pseudo-Hermitian limit.

\subsection{Adiabatic theorem: Consistently defined covariant Berry connection in pseudo-Hermitian limit}

In the previous subsection it was shown that, for a generic non-Hermitian Hamiltonian, the natural geometric structure is the AA holonomy associated with cyclic evolution in projective Hilbert space. A reduction to the conventional Berry framework becomes possible only under additional structural constraints.

Specifically, when the metric does not exhibit explicit time dependence, the evolution admits a consistent adiabatic description in terms of instantaneous eigenstates, allowing one to define a Berry connection on the parameter-space base manifold ${\cal M}_\lambda$, in analogy with standard Hermitian quantum mechanics.

This situation is realized in pseudo-Hermitian systems, where the Hamiltonian possesses a real spectrum and the Hermitizing transformation can be chosen such that the metric has no explicit time dependence (while retaining an implicit dependence through $\boldsymbol{\lambda}(t)$). In this case, \eqref{HH} reduces to
\begin{equation}
    H^H = S H S^{-1}, \qquad \Gamma^t = 0.
\end{equation}

Accordingly, to recover Berry holonomy in the conventional sense, i.e., as a holonomy generated by a connection on the eigenbundle over the parameter manifold ${\cal M}_\lambda$, we consider a slowly varying Hermitian Hamiltonian satisfying the instantaneous eigenvalue equation
\begin{equation}
H^H(t)
|\phi_n^H(t)\rangle
=
E_n^H(t)
|\phi_n^H(t)\rangle,
\end{equation}
with $E_n^H(t)\in\mathbb R$ being a non-degenerate instantaneous eigenenergy.

By expanding an evolving state $|\phi^H(t)\rangle$ as
\begin{equation}
|\phi^H(t)\rangle
=
\sum_n
a_n(t)
\exp\left[
-i
\int_0^t
E_n^H(t')dt'
\right]
|\phi_n^H(t)\rangle,
\end{equation}
and, using the Schr\"odinger equation in~\eqref{SEH}, we obtain
\begin{eqnarray}
   \dot a_m
   &=&
   -a_m
   \langle\phi_m^H|
   \dot\phi_m^H\rangle
   \nonumber\\
   &&
   -
   \sum_{n\neq m}
   a_n
   \frac{
   \langle
   \phi_m^H|
   \dot H^H
   |\phi_n^H
   \rangle
   }{
   E_n^H-E_m^H
   }
   \exp\!\left[
   i
   \int_0^t
   (E_m^H-E_n^H)
   dt'
   \right]. \nonumber \\
\end{eqnarray}

The adiabatic regime is characterized by~\cite{Berry1984_BP}
\begin{equation}
\sum_{n\neq m}
\left|
\frac{
\langle
\phi_m^H|
\dot H^H
|\phi_n^H
\rangle
}{
(E_n^H-E_m^H)^2
}
\right|
\ll1,
\end{equation}
ensuring that transitions between instantaneous eigenstates are suppressed.

Assuming $a_n(0)=\delta_{mn}$ and an adiabatic evolution along a closed loop in parameter space of period $T$, with $\boldsymbol{\lambda}(T)\equiv\boldsymbol{\lambda}(0)$, and with no spectral singularities encountered, the state evolves as
%Assuming $a_n(0)=\delta_{mn}$ and an adiabatic evolution along a closed loop in parameter space of period $T$, with $\boldsymbol{\lambda}(T)\equiv\boldsymbol{\lambda}(0)$ and no spectral singularities encountered, the state evolves as
\begin{equation}
    |\phi_m^H(T)\rangle=e^{\displaystyle-i\int_0^T E_m^H(t)\,{\rm d}t}\ e^{\displaystyle i\oint { A}_m^H  {\rm d}{\boldsymbol{\lambda}}}\,|\phi_m^H(0)\rangle,
\end{equation}
where $A_m^H = i\langle\phi_m^H|\nabla_\lambda|\phi_m^H\rangle$ is the Berry connection.

Again, transforming back to the non-Hermitian frame gives
\begin{equation}\label{Revol}
|\psi_m^R(T)\rangle
=
S_T^{-1}S_0\,
e^{\displaystyle-i\int_0^T E_m^H(t)\,dt}
\,
e^{\displaystyle i\oint A_m^H\, d\boldsymbol{\lambda}}
|\psi_m^R(0)\rangle.
\end{equation}
Equation~(\ref{Revol}) shows that, in this pseudo-Hermitian adiabatic regime, the geometric contribution to the evolution is fully captured by the Berry phase
\[
\gamma_B = \oint A_m^H\, d\boldsymbol{\lambda}.
\]
Similarly to~\eqref{Revol_AA}, the quantity $S_T^{-1}S_0$ ensures that the inner product $\langle\psi_m^R(0)|\eta(0,T)|\psi_m^R(T)\rangle\in \rm U(1)$.

Consequently, for an arbitrary (possibly degenerate) right eigenspace, the resulting Berry connection assumes a covariant Hermitian form
\begin{equation}\label{Ac}
    {\mathbb A}^{\lambda}_{mn}={A}^{\lambda H}_{mn}
    =i\langle \phi^H_m|\partial_\lambda|\phi^H_n\rangle
    =i\langle \psi^R_m|\eta D_\lambda|\psi^R_n\rangle,
\end{equation}
or, equivalently,
\begin{equation}\label{AG}
    {\mathbb A}^\lambda_{mn}
    ={\cal A}^{LR}_{\lambda,mn}
    +i\langle \psi^L_m|\Gamma^\lambda|\psi^R_n\rangle,
\end{equation}
where 
\begin{equation}\label{Ain}
    {\cal A}^{LR}_{\lambda,mn}=\langle\psi^L_{m}|\partial_\lambda|\psi^R_n\rangle,
\end{equation}
is the conventional left-right Berry connection~\cite{Silberstein2020,Singhal2023,Ozawa2025b,Montag2026}, and $D_\lambda$ with $\Gamma^\lambda$ are given in~\eqref{D}.

In the Hermitian limit, the covariant Berry connection (CBC) $\mathbb A^\lambda_{mn}$ reduces to the standard Berry connection, since $\eta\to I$ and $D_\lambda\to\partial_\lambda$. Likewise, when the metric is constant, i.e., $\Gamma^\lambda=0$, the CBC reduces to the usual left-right non-Hermitian Berry connection given in~\eqref{Ain}. In this sense, the standard left--right connection ${\cal A}^{LR}$ is appropriate in the genuine quantum regime only when the Hilbert-space metric is {\it constant} in parameter space.

In direct analogy with \eqref{HH}, where the time-like connection compensates for the non-Hermitian contribution to the spectrum, the parameter-space metric-compatible connection $\Gamma^\lambda$ compensates for the contribution to the Berry connection arising from the rotations and rescaling of the metric, thus isolating the geometry associated with the physical, norm-preserving, eigenspace of the NHH. 
This demonstrates that the covariant Berry connection is not an arbitrary extension, but rather the unique geometric object {\it physically consistent with the norm-preserving quantum evolution of a state}. 

In this context, it is also instructive to compare with an alternative definition within the metric formalism proposed in Ref.~\cite{Ju2024}, where the Berry connection is obtained by imposing parallel transport of the right eigenvectors in the $\eta$-deformed Hilbert space, i.e., $D_\lambda|\psi\rangle=0$. While this approach is mathematically consistent and provides valuable insight, it relies on an additional constraint that is not dictated by the underlying quantum dynamics. Namely, under this assumption, the metric-compatible connection $\Gamma$
 becomes equivalent to the conventional left-right Berry connection 
${\cal A}^{LR}$.
Conceptually, however, the two are defined on distinct structures:  $\Gamma^\lambda$
 accounts for the geometry of the $\eta$-deformed Hilbert space, whereas ${\cal A}^{LR}$ encodes the geometry and topology of the Hamiltonian eigenbundle. Moreover, in the Hermitian limit, $\Gamma^\lambda\to 0$, this would imply that the Berry connection can always be set to zero, thus leading to inconsistencies.

 On the other hand, the metric-dependent Berry connection introduced in Ref.~\cite{Zhang2019top_obst} is only applicable for pseudo-Hermitian systems with a specific form of metric-compatible connection $\Gamma=\eta^{-1}d\eta/2$. This corresponds to a particular choice of the connection in~\eqref{Gamma_def} and effectively assumes that the eigenvector space of $\eta$ is independent of the system parameters. As a result, this construction does not exhibit full covariance~\cite{Mostafazadah2018,arkhipov2026a}.

From the discussion above it becomes evident that a Berry connection in the conventional sense, i.e., understood as a geometric holonomy associated with closed trajectories in the parameter manifold ${\cal M}_\lambda$, can be consistently defined only in the pseudo-Hermitian limit. Consequently, in the present work we primarily focus on this regime in order to formulate the Berry connection for non-Hermitian systems within this conventional geometric framework. 
Importantly, even within this restricted setting, as we show below, the covariant formulation of the Berry connection reveals a nontrivial departure from the standard biorthogonal framework in the definition of holonomies. 
This highlights that the pseudo-Hermitian limit does not merely recover the standard biorthogonal Berry phase, but instead provides a covariant geometric structure in which the metric degrees of freedom actively participate in defining the Berry connection and its associated holonomy.

\section{Gauge transformation of the covariant Berry connection} \color{black}
In the following sections, we restrict our attention to the pseudo-Hermitian limit, in which the studied geometric structure reduces to a conventional parameter-space eigenbundle over ${\cal M}_\lambda$. We then focus on the gauge structure and curvature associated with the parameter-space CBC ${\mathbb A}^\lambda$ introduced in~\eqref{Ac}.

{Here we discuss the transformation properties of the CBC under general ${\rm GL}(N,\mathbb{C})$ gauge transformations.} Such transformations arise either from the intrinsic biorthogonal gauge freedom of left and right eigenvectors or from transition functions relating local trivializations across overlapping patches when the CBC is defined over a nontrivial parameter-space bundle.

Let $|R\rangle$ ($\langle L|$) denote the matrices consisting of columns (rows) of right (left) $N$ degenerate eigenvectors.
One can verify that, for such an $N$-dimensional eigenvector frame of the NHH $H$, an arbitrary frame transformation $T\in{\rm GL}(N,\mathbb{C})$:
$|R'\rangle = |R\rangle T$ and  $\langle L'| = T^{-1}\langle L|$ 
induces the following {\it affine} gauge transformation of the CBC:
\begin{equation}\label{Atilde}
    {\mathbb A}'^{\lambda}
= T^{-1} \tilde{\mathbb A}^{\lambda} T+iT^{-1}\partial_\lambda T, \quad \tilde{\mathbb A}^{\lambda}={\mathbb A}^\lambda+\Xi^\lambda,
\end{equation}
where
\begin{equation}\label{Xi}
\Xi^\lambda
= i\langle R|\eta\left(\Gamma'^\lambda
- \Gamma^\lambda\right)|R\rangle=i\langle R|\eta \left(T\partial_\lambda T^{-1}\right)|R\rangle,
\end{equation}
with $
\Gamma'^{\lambda}=\Gamma^{\lambda}+T\partial_\lambda T^{-1}$ being the transformed metric-compatible connection under the frame change. This follows from the covariance of the covariant derivative under the frame change, i.e., $D'_\lambda|R'\rangle=(D_\lambda|R\rangle)T$.

Equation~(\ref{Atilde}) shows that the CBC does not transform as a standard gauge potential, but rather as an affine connection built from the shifted quantity $\tilde{\mathbb A}^\lambda={\mathbb A}^\lambda+\Xi^\lambda$. The additional term $\Xi^\lambda$ originates from the transformation of the metric-compatible connection $\Gamma^\lambda$ and therefore encodes the change of the Hilbert-space metric under ${\rm GL}(N,\mathbb C)$ frame transformations. In this sense, $\Xi^\lambda$ is not an independent gauge field, but a metric-induced correction that ensures covariance of the CBC in a $\eta$-deformed Hilbert space. 
More specifically, $\Xi^\lambda$ stems from the non-unitary (positive-definite) part $P$ in the polar decomposition of $T=UP$. In other words, the affine term compensates for the local rescaling of eigenvectors induced by $P$.
For a purely unitary transformation $T=U$, the $(\Gamma'-\Gamma)$ is a pure gauge. This means that together with the inner unitary gauge freedom of the metric connection, one can always set $\Xi^\lambda=0$, and thus $\mathbb A_\lambda$ transforms as an ordinary ${\rm U}(N)$ gauge potential. 
Evidently, in the Hermitian limit $\Gamma^\lambda\to0$, $\Xi^\lambda\to0$, the \eqref{Atilde} reduces the standard Berry connection transformation law~\cite{Wilczek1984}.

To recap, the transformation law, in~\eqref{Atilde}, separates two distinct geometric contributions: the intrinsic geometry of the Hamiltonian eigenbundle, captured by ${\mathbb A}^\lambda$, and the geometry associated with the parameter-dependent metric, encoded in $\Xi^\lambda$. That is, the CBC effectively isolates the physical, norm-preserving eigenspace geometry by consistently eliminating metric-induced effects via the affine term.
This is to compare with the conventional biorthogonal gauge transformation law 
where no such compensation is provided, and where the metric and intrinsic eigenspace geometry are becoming mixed under ${\rm GL}(N,\mathbb C)$ frame transformations.

\section{Berry curvature}
Evidently, the Berry curvature associated with the CBC has exactly the same algebraic structure as in Hermitian systems. In the general differential form and component notation, respectively, it reads~\cite{Xiao2010}
\begin{equation}\label{F}
F={\rm d}{\mathbb A}-i\mathbb{A}\wedge \mathbb{A}, \quad
F_{\nu\mu}=\partial_\nu {\mathbb A}^\mu-\partial_\mu {\mathbb A}^\nu-i[{\mathbb A}^\nu,{\mathbb A}^\mu],
\end{equation}
where ${\rm d}$ denotes the ordinary exterior derivative acting on the connection one-form, and $\wedge$ is the graded wedge product for matrix-valued differential forms~\cite{GriffithsBook}.
This immediately implies that, given the affine transformation law \eqref{Atilde}, the Berry curvature transforms accordingly under a general frame change $T\in{\rm GL}(N,\mathbb C)$ as
\begin{equation}\label{Ftilde}
F'=T^{-1}\tilde FT, \qquad
\tilde F={\rm d}\tilde {\mathbb A}-i\tilde{\mathbb{A}}\wedge \tilde{\mathbb{A}},%F=F+F_\Xi,
\end{equation}
with $\tilde{\mathbb{A}}$ given in \eqref{Atilde}.
Importantly, the affine curvature $\tilde F$ represents the genuine gauge-covariant curvature of the eigenbundle associated with norm-preserving state vectors under frame transformations.

Because both the CBC and curvature defined in Eqs.~(\ref{Atilde}) and (\ref{Ftilde}) are uniquely specified, the corresponding topological invariants, such as Chern numbers, are likewise uniquely defined. This contrasts with the four definitions of Chern numbers for Abelian eigenbundles discussed in Ref.~\cite{Shen2018}, which, although distinct, yield the same invariant. Within the covariant framework, however, the resulting topology of the physical (norm-preserving) eigenbundle may differ from the conventional biorthogonal one~\cite{Shen2018}, since in the latter, the non-Hermitian curvature can arise purely from the metric rather than from intrinsic geometry of the eigenspace (see Sec.~\ref{Examples}), and therefore can be fictitious. Notably, all contributions involving the affine term $\Xi$ in \eqref{Ftilde} are topologically trivial: they either vanish or reduce to exact differential forms, since they are generated pure gauge terms, and therefore do not contribute to Chern integrals over closed manifolds.

\section{Examples}\label{Examples}
{Let us illustrate the proposed framework with a couple of concrete examples, showcasing how the CBC effectively handles the biorthogonal gauge freedom and in parallel highlighting the fundamental difference between the conventional left-right and covariant Berry connection formalisms.} 
\subsection{Abelian covariant Berry connection}
We first consider a pseudo-Hermitian Hamiltonian of the form:
\begin{equation}\label{Hph}
    H^{\rm pH}=t\sigma_z+ix\sigma_y-iy\sigma_x+a_0I, \quad t,x,y,a_0\in {\mathbb R},
\end{equation}
where $\sigma_i$ are Pauli matrices.
This two-level Hamiltonian can, for instance, effectively describe topological magnons on a honeycomb lattice~\cite{Cheng2016,Berakdar2026}. The same model was recently analyzed in Ref.~\cite{Deng2025}, in the context of emerging synthetic electromagnetic fields.  

 The eigenvalues of the Hamiltonian are 
 \begin{equation}
     E_\pm=a_0\pm l, \quad \text{with}\quad  l=\sqrt{t^2-x^2-y^2},
 \end{equation}
implying that the condition $l>0$ defines an exact pseudo-Hermitian phase where the eigenspectrum is real. Following Ref.~\cite{Deng2025}, for simplicity, we focus on a single isolated energy band with $E_-$. Through the hyperbolic parameterization: 
\begin{eqnarray}
    t=l\cosh\xi, \quad x=l\sinh\xi\cos\lambda, \quad y=l\sinh\xi\sin\lambda,
\end{eqnarray}
the eigenvectors take the form on the reduced 2-dimensional  parameter manifold $(\lambda,\xi)$~\cite{Deng2025}:
 %. Specifically, for the single energy-band $E_$, they read~\cite{Deng2025}:
\begin{eqnarray}\label{psi1}
    |\psi^R_-\rangle=\begin{pmatrix}-e^{-i\lambda}\sinh\frac{\xi}{2}\\ 
    \cosh\frac{\xi}{2}\end{pmatrix},\quad |\psi^L_-\rangle=\begin{pmatrix}e^{-i\lambda}\sinh\frac{\xi}{2}\\\cosh\frac{\xi}{2}\end{pmatrix}.
\end{eqnarray}
By fixing $E_-=const\neq a_0$, and exploiting Eqs.~(\ref{Ain}) and (\ref{psi1}), the corresponding conventional left-right Berry connection reads
\begin{eqnarray}\label{Alr}
    A^{LR}_{-}=i\langle\psi^L_-|{\rm d}|\psi^R_-\rangle=-\sinh^2\frac{\xi}{2}d\lambda.
\end{eqnarray}
Now, using the biorthogonal gauge freedom for the eigenvectors in \eqref{psi1}, one may rescale both of them by a normalized factor
\begin{equation}
    N=\sqrt{\sinh^2\frac{\xi}{2}+\cosh^2\frac{\xi}{2}},
\end{equation}
(the normalization factor $N$ is chosen such that $\langle\psi'^R_i|\psi'^R_i\rangle=1$ for $i=\pm$), namely 
\begin{equation}
    |\psi'^R_i\rangle=N^{-1}|\psi^R_i\rangle, \quad |\psi'^L_i\rangle=N|\psi^L_i\rangle,
\end{equation}
such that 
\begin{equation}
    \langle\psi'^L_i|\psi'^R_i\rangle=\langle\psi^L_i|\psi^R_i\rangle=1, \quad i=\pm.
\end{equation}
That is, in this particular case the frame transformation is 
\begin{equation}
    T=N^{-1}I_2\in {\rm GL}(2,\mathbb R).
\end{equation}
%$T=N^{-1}I_2\in {\rm GL}(2,\mathbb R)$. 
Because of this the conventional left-right Berry connection transforms as
\begin{equation}\label{A'lr}
A'^{LR}_-= A^{LR}_{-}-\frac{i}{2}(\tanh\xi) d\xi.   
\end{equation}
Evidently, after such a rescaling the left–right Berry connection acquires an additional imaginary term. As a consequence, along a given open-path trajectory the Berry phase develops a dissipative component $\gamma_d\equiv\int (\tanh\xi)\mathrm d\xi$. This shows that within the conventional non-Hermitian formalism the biorthogonal gauge freedom alone can induce a geometric dissipation of a state along open paths in the system parameter space, though the energy remains real.

Now we consider the CBC. 
For that purpose, one must first find the connection 
\begin{equation}
    \Gamma=S^{-1}dS=\Gamma^\lambda d\lambda+\Gamma^\xi d\xi,
\end{equation}
according to Eqs.~(\ref{D}) and (\ref{Ac}). The components of the connection are found as  (see Appendix~\ref{AA} for details): 
\begin{equation}\label{Gamma_comp}
    \Gamma^\lambda=\begin{pmatrix}
        i & 0 \\
        0 & 0
    \end{pmatrix}, \quad \Gamma^\xi=\dfrac{1}{2}\begin{pmatrix}
        0 & e^{-i\lambda} \\
        e^{i\lambda} & 0
    \end{pmatrix}.
\end{equation}
The CBC in \eqref{Ac} then reads (see Appendix~\ref{AB}):
\begin{equation}\label{A-}
    {\mathbb A}_-=0.
\end{equation}
That is, under the wave-function norm–preserving condition, the CBC vanishes identically. As a result, no geometric phase is accumulated when the state undergoes a cyclic evolution. This is in contrast to the conventional left–right Berry connection in~\eqref{Alr}, which generally depends on $\xi$. {In other words, the nonzero left-right Berry connection in \eqref{Alr} arises solely from the metric, rather than from the intrinsic geometry of the Hamiltonian eigenspace.} 

Crucially, under the rescaling of the eigenvectors by the factor $N$, the associated CBC, according to \eqref{Atilde}, remains unchanged, i.e., 
\begin{equation}\label{A'-}
    {\mathbb A}'_- = 0, 
\end{equation}
(see Appendix~\ref{AD} for details).
This demonstrates that the CBC is insensitive to such length-changing transformations of the eigenvectors. Indeed, the affine component of the CBC in \eqref{Xi} (see Appendix~\ref{AD}), 
\begin{equation}
    \Xi=\frac{i}{2}\tanh\xi d\xi,
\end{equation}
exactly cancels the arising imaginary contribution in the left-right Berry connection in \eqref{A'lr}, thus ensuring norm preservation of the quantum state under adiabatic evolution.

In light of the above, it is clear that, for this analyzed model, no Berry curvature  is generated within the CBC formalism, i.e., 
\begin{equation}
    F=0,
\end{equation}
according to~\eqref{F}, in the entire $(\lambda,\xi)$-space. This stands in contrast to the nonzero curvature $F=-i(\sinh\xi)\sigma_y/2$~\cite{Deng2025} obtained from the left-right Berry connection in~\eqref{Alr}.

These findings can be additionally understood from the perspective of the Hermitizing maps $S$ and $S'$ associated with the original and rescaled frames. As shown in Appendix~\ref{AC}, the corresponding eigenvectors of the Hermitian Hamiltonians,
\begin{equation}
    H^{H}=S H^{\rm pH} S^{-1}\equiv l\sigma_x,
\end{equation}
and 
\begin{equation}
    H'^{H}=S' H^{\rm pH} S'^{-1}\equiv l\cosh\xi\sigma_x,
\end{equation}
namely
\begin{equation}
    |\phi^H_-\rangle = S|\psi^R_-\rangle, \quad \text{and} \quad |\phi'^H_-\rangle = S'|\psi'^R_-\rangle,
\end{equation}
are equivalent up to an arbitrary phase, with the globally smooth $S$ and $S'$. Explicitly, one finds
\begin{eqnarray}
|\phi'^H_-\rangle \equiv |\phi^H_-\rangle \equiv\bigl[1,1\bigr]^T.
\end{eqnarray}
That is, the eigenvectors associated with the Hermitian Hamiltonian are fixed throughout the entire $(\xi,\lambda)$-space, and the corresponding Berry connections, therefore, vanish identically in this region. In other words, the apparent rotations of the right eigenvectors in the non-Hermitian frame arise solely from the metric: the metric eigenvectors, in addition to the rescaling determined by $\xi$, undergo parameter-dependent rotations governed by $\lambda$ (see Appendix~\ref{AA}). However, the intrinsic geometry of the Hamiltonian eigenspace is trivial.

\subsection{Non-Abelian covariant Berry connection}
Let us consider another minimal, analytically tractable, example of a pseudo-Hermitian system, now featuring an exceptional point (EP) in its spectrum. The Hamiltonian reads
\begin{eqnarray}\label{H2}
    H_0=(\epsilon-i\Delta)\sigma_z+(k+i\chi)\sigma_x.
\end{eqnarray}
Such a model describes, for instance, two coupled dissipative optical cavities (in the mode representation), where $\Delta$ ($-\Delta$) denotes the gain (loss) rate and $\epsilon$ the frequency detuning. The parameters $k$ and $\chi$ correspond to coherent and incoherent coupling strengths, respectively.

Introducing the hyperbolic parameterization~\cite{arkhipov2024a}
\begin{eqnarray}\label{sys_param}
    &\Delta = \alpha\sinh\phi_i\sin\phi_r, \quad \epsilon =\alpha\cosh\phi_i\cos\phi_r,& \nonumber \\
    &k = \alpha\cosh\phi_i\sin\phi_r, \quad \chi =\alpha\sinh\phi_i\cos\phi_r,&
\end{eqnarray}
where $\phi=\phi_r+i\phi_i=\arctan(z^{-1})\in{\mathbb C}$ with $z=x+iy$, $x,y\in {\mathbb R}$, and $\alpha=x\sinh\phi_i/\sin\phi_r\in{\mathbb R}$, the Hamiltonian in \eqref{H2} takes the form
\begin{eqnarray}\label{H}
    H = \alpha\begin{pmatrix}
        \cos\phi & \sin\phi \\
        \sin\phi & -\cos\phi
    \end{pmatrix}, \quad E_{1,2}=\mp\alpha.
\end{eqnarray}
The corresponding right and left eigenvectors of $H$ are
\begin{eqnarray}\label{psi3}
  & |r_1\rangle\equiv
        \left[-\sin\frac{\phi}{2},
        \cos\frac{\phi}{2}\right]^T, \quad
|l_1\rangle\equiv
        \left[-\sin\dfrac{\phi^*}{2},
        \cos\dfrac{\phi^*}{2}\right]^T, & \nonumber \\
        & |r_2\rangle\equiv
        \left[\cos\dfrac{\phi}{2},
        \sin\dfrac{\phi}{2}\right]^T, 
           \quad
    |l_2\rangle\equiv
        \left[\cos\dfrac{\phi^*}{2},
        \sin\dfrac{\phi^*}{2}\right]^T. &
\end{eqnarray}

The spectrum exhibits two EPs in the parameter space at $x=0,y=\pm1$, i.e.,  at $z=\pm i$ (see Fig.~2 in Ref.~\cite{arkhipov2024a}). At these points, two real-valued Riemann energy sheets intersect, forming branch points, i.e., EPs.

The presence of such branch points implies that eigenstates can undergo permutation upon encircling either of them. To capture this nontrivial holonomy, one must therefore employ the non-Abelian Berry connection. We begin by evaluating the conventional left–right Berry connection. Combining Eqs.~(\ref{Ain}) and (\ref{psi3}), we obtain
\begin{eqnarray}\label{Alr2}
    {\cal A}^{LR}=\frac{\sigma_y}{2(1+z^2)}\,{\rm d}z, \quad \left({\cal A}^{LR}\right)^\dagger\neq {\cal A}^{LR}.
\end{eqnarray}
Given the form of \eqref{Alr2}, the holonomy associated with a loop encircling the EP at $z=i$ is readily evaluated as
\begin{eqnarray}\label{hol}
    U_{\rm hol}={\cal P}\exp\left(\frac{i}{2}\oint\frac{\sigma_y{\rm d}z}{1+z^2}\right)=i\sigma_y\in {\rm SU}(2).
\end{eqnarray}
This result reflects the characteristic behavior of an EP: after a single encircling, the eigenstates are exchanged, with one of them acquiring an additional phase of $\pi$. 

Notably, using the conventional biorthogonal formalism, the resulting holonomy is revealed to be unitary, despite ${\cal A}^{LR}$ is non-Hermitian, contrasting with typical non-unitary holonomies in non-Hermitian systems when encircling an EP~\cite{Uzdin2011,Graefe2013,Shan2024}.
This also indicates that this unitary holonomy can be potentially observed in {\it (semi)classical} systems described by the NHH $H$, where consistent accounting for state-norm preservation is unnecessary.

We now examine whether this holonomy reflects intrinsic geometry of the Hamiltonian physical eigenspace or merely a metric-induced artifact in the {\it genuine quantum regime}. To this end, we construct the metric $\eta$. Using Eqs.~(\ref{eta_dec}) and (\ref{psi3}), we find
\begin{equation}\label{eta2}
    \eta=I_2\cosh\theta-\sigma_y\sinh\theta, \quad \theta={\rm Im}\phi=\frac{1}{4}\ln\frac{x^2+(y-1)^2}{x^2+(y+1)^2}. \nonumber
\end{equation}
Diagonalizing $\eta$ yields 
\begin{equation}
    D={\rm diag}[e^{-\theta/2},e^{\theta/2}], \quad U=\dfrac{1}{\sqrt{2}}\begin{pmatrix}
    -i & i \\
    1 & 1
\end{pmatrix}.
\end{equation}
The corresponding vielbein $S$ then reads
\begin{equation}
     S=\dfrac{1}{\sqrt{2}}\begin{pmatrix}
        ie^{-\theta/2} & e^{-\theta/2} \\
        -i e^{\theta/2} & e^{\theta/2}
    \end{pmatrix}.
\end{equation}
Accordingly, the metric connection one-form is 
\begin{equation}
    \Gamma=-\sigma_y\,{\rm d}\theta/2.
\end{equation}
 Consequently, the CBC, in \eqref{AG}, attains the following Hermitian form
\begin{equation}\label{A3}
    {\mathbb A}={\cal A}^{LR}+i\langle L|\Gamma|R\rangle=\frac{\sigma_y}{2}\left(\frac{{\rm d}z}{1+z^2}+i\,{\rm d}\theta\right), \quad {\mathbb A}^\dagger={\mathbb A}, \nonumber
\end{equation}
where $\langle L|$ ($|R\rangle$) is a matrix consisting of rows (columns) of the left (right) eigenvectors in~\eqref{psi3}. Evidently ${\mathbb A}\neq {\cal A}^{LR}$.
The metric correction in the CBC is thus encoded in the term $i\sigma_y{\rm d}\theta/2$. However, for a closed loop encircling the EP, one has
\begin{equation}
    \oint {\rm d}\theta=0,
\end{equation}
since $\theta$ is single-valued along the cycle. Therefore, this metric compensation does not contribute to the holonomy in \eqref{hol}, and both the left–right and covariant formulations yield identical results. In other words, even
though ${\cal A}^{LR}$ mixes intrinsic geometry and metric, here, these metric-induced effects are effectively canceled out in the holonomy.

Moreover, although the two Berry connections differ locally, both give the same curvature, $F=0$, away from the EP, since they are effectively determined by a single differential ${\rm d}\phi$. As a result, no Chern number can be associated with the EP. Instead, here the topology is encoded in the loop-shape-independent holonomy: the exchange of eigenstates upon a single encircling defines a $\mathbb{Z}_2$ topological invariant, reflecting the branch-point structure of the spectrum. 

Together with the example in the main text, where the conventional biorthogonal formalism produces purely metric-induced (spurious) holonomy, this clearly demonstrates that: (i) the CBC eliminates unphysical geometric artifacts when present, and (ii) it fully retains nontrivial geometry and topology when they are intrinsic, therefore providing a physically consistent geometric framework in  quantum  pseudo-Hermitian systems.

\section{Conclusions}

In this work, we demonstrated that, at the quantum level, when the dynamics is described by continuous non-unitary evolution, the Berry connection in the conventional sense, namely, as a geometric connection defined over closed loops in the parameter manifold, can be consistently formulated only for pseudo-Hermitian Hamiltonians characterized by real spectra, i.e., whose associated Hilbert space metric structure does not explicitly depend on time. 
For such systems, we constructed a covariant Berry connection that faithfully characterizes the geometry and topology of pseudo-Hermitian eigenbundles, while resolving gauge ambiguities inherent to the standard biorthogonal formulation.

The proposed covariant framework consistently incorporates general ${\rm GL}(N,\mathbb C)$ frame transformations and disentangles the intrinsic geometry of the physical eigenspace from contributions induced by the parameter-dependent Hilbert-space metric. From this perspective, the geometry captured by the conventional biorthogonal approach with {ad hoc}  renormalization generally {\it conflates} intrinsic and metric-induced contributions, and may therefore {\it overestimate} the geometry of the Hamiltonian eigenbundle, leading to the emergence of fictitious holonomies and Berry curvatures in the genuine quantum regime. Accordingly, the non-Hermitian Berry phases and synthetic gauge fields reported in Ref.~\cite{Deng2025} and related works can be reinterpreted as manifestations of the metric structure in regimes where state-norm preservation is not essential. In contrast, within the covariant (norm-preserving) framework, such geometric phases and synthetic fields may vanish, revealing their metric origin rather than intrinsic properties of the physical Hamiltonian eigenspace.

More generally, we have argued that for non-Hermitian Hamiltonians with genuinely complex spectra, the conventional notion of Berry holonomy associated with closed trajectories in parameter space is not, in general, sufficient. In such cases, the natural geometric structure is instead described within the Aharonov--Anandan framework, since the evolution is generically closed only in projective Hilbert space rather than in the parameter manifold. From this viewpoint, the appropriate geometric object is the Aharonov--Anandan holonomy associated with cyclic evolution of the quantum ray, while the conventional Berry phase arises as a special pseudo-Hermitian limit of this more general projective-space construction.

Overall, this work establishes a covariant geometric framework for non-Hermitian quantum systems that consistently distinguishes physical geometric content from metric-induced artifacts, and clarifies the precise domain of validity of the Berry connection and associated Berry holonomy in non-Hermitian settings.

\begin{acknowledgments}
The author would like to thank Chia-Yi Ju for insightful discussions. The author also acknowledges support from the Air Force Office of Scientific Research (AFOSR) Award No. FA8655-24-1-7376, 
from the Grant Agency of the Czech Republic (Project No. 25-15775S).
%and from the Ministry of Education, Youth and Sports of the Czech Republic Grant OP JAC No. CZ.02.01.01/00/23\_021/0008790.
\end{acknowledgments}

%\clearpage
%\newpage
\appendix
%\clearpage
%\renewcommand{\appendixname}{End Matter}
%\begin{widetext}
\section{Calculating metric-compatible connection $\Gamma$ for the Hilbert space of pseudo-Hermitian Hamiltonian in \eqref{Hph}}\label{AA}
Here we elaborate on the explicit forms of the metric-compatible connections $\Gamma$ and $\Gamma'$ for the Hilbert space of the pseudo-Hermitian Hamiltonian in~\eqref{Hph} spanned by the original eigenvectors in \eqref{psi1},  and  by the eigenvectors rescaled by the real factor $N$ introduced below~\eqref{Alr}, respectively.

Let us start with the connection
\begin{equation}\label{Gamma}
    \Gamma = S^{-1}\mathrm d S=\Gamma^\lambda d\lambda+\Gamma^\xi d\xi.
\end{equation}
 To construct it explicitly, one first needs to determine the Hermitizing map $S$. This map can be obtained using the standard decomposition~\cite{arkhipov2026a},
\begin{equation}\label{Sdec}
S = \sqrt{D}U^\dagger ,
\end{equation}
where $D$ and $U$ are diagonal and unitary matrices, respectively, arising from the spectral decomposition of the metric $\eta = S^\dagger S$, which is diagonalized as $\eta = U D U^\dagger$ (away from the exceptional points). The metric is first attained via the expression 
\begin{eqnarray}\label{eta_dec}
    \eta=\sum\limits_{i=\pm}|\psi^L_i\rangle\langle\psi^L_i|.
\end{eqnarray}
Now combining Eqs.~(\ref{eta_dec}) and (\ref{psi1}) one finds
\begin{equation}
    \eta=\begin{pmatrix}
        \cosh\xi &e^{-i\lambda}\sinh\xi \\
        e^{i\lambda}\sinh\xi & \cosh\xi
    \end{pmatrix}.
\end{equation}
The eigenvalues and eigenvectors of the metric are thus read
\begin{equation}
    D=\begin{pmatrix}
        e^{-\xi} & 0\\
        0 & e^{\xi}
    \end{pmatrix}, \quad U = \frac{1}{\sqrt{2}}\begin{pmatrix}
        -e^{-i\lambda} & e^{i\lambda} \\
        1 & 1
    \end{pmatrix}.
\end{equation}
As a result, the Hermitizing map attains the form (up to an arbitrary U(2) gauge):
\begin{eqnarray}\label{S_expl}
    S = \frac{1}{\sqrt{2}}\begin{pmatrix}
        -e^{-\frac{\xi}{2}+i\lambda} & e^{-\frac{\xi}{2}} \\
        e^{\frac{\xi}{2}+i\lambda} & e^{\frac{\xi}{2}}
    \end{pmatrix}.
\end{eqnarray}
Utilizing Eqs.~(\ref{Gamma}) and (\ref{S_expl}), one then arrives at the expressions for $\Gamma^{\lambda,\xi}$ in \eqref{Gamma_comp}.
Note that the obtained map $S$ is continuous everywhere in the system parameter space. 

\renewcommand{\theequation}{B\arabic{equation}}
\setcounter{equation}{0}
\section{Calculating the covariant Berry connection in \eqref{A-}}\label{AB}
According to \eqref{Ac}, the CBC corresponding to the eigenenergy $E_-$ is defined as
\begin{equation}
    {\mathbb A}_-=i\langle\psi^R_-|\eta D_\lambda|\psi^R_-\rangle d \lambda+i\langle\psi^R_-|\eta D_\xi|\psi^R_-\rangle d \xi.
\end{equation}
Exploiting the expressions for $\Gamma^{\lambda,\xi}$ in the main text, one obtains the CBC components as follows
\begin{eqnarray}
    {\mathbb A}^\lambda=&&i\langle\psi^L_-|\partial_\lambda|\psi^R_-\rangle+i\langle\psi^L_-|\Gamma^\lambda|\psi^R_-\rangle \nonumber \\
    &&=(-\sinh^2\frac{\xi}{2}+\sinh^2\frac{\xi}{2})d\lambda=0,
\end{eqnarray}
and 
\begin{eqnarray}
    {\mathbb A}^\xi=i\langle\psi^L_-|\partial_\xi|\psi^R_-\rangle+i\langle\psi^L_-|\Gamma^\xi|\psi^R_-\rangle=(0+0)d\xi=0.
\end{eqnarray}
Consequently, one arrives at~\eqref{A-} with $\mathbb A=0$.

\renewcommand{\theequation}{C\arabic{equation}}
\setcounter{equation}{0}
\section{Transformed covariant Berry connection}\label{AD}
Let us now explicitly compute the transformed CBC $\mathbb A'$, below \eqref{A-}, using the modified metric-compatible connection on the Hilbert space $\Gamma' = S'^{-1}\mathrm d S'$. Combining this expression with~(\ref{S'}) and for $\Gamma^{\lambda,\xi}$ in \eqref{Gamma_comp}, one finds the corresponding components
\begin{equation}\label{G'}
     \Gamma'^\lambda= \Gamma^\lambda=\begin{pmatrix}
        i & 0 \\
        0 & 0
    \end{pmatrix}, \quad \Gamma'^\xi=\frac{1}{2}\begin{pmatrix}
        \tanh\xi & e^{-i\lambda} \\
        e^{i\lambda} & \tanh\xi
    \end{pmatrix}.
\end{equation}
Substituting~\eqref{G'} into~\eqref{Ac}, one immediately obtains $\mathbb A' = 0$.
Alternatively, one can find $\Gamma'$ straightforwardly from the relation $\Gamma'^{\lambda}=\Gamma^{\lambda}+T\partial_\lambda T^{-1}$, with $T=N^{-1}I_2 \in {\rm GL}(2,\mathbb R)$ and $N=\sqrt{\sinh^2\frac{\xi}{2}+\cosh^2\frac{\xi}{2}}$, according to the main text.

It is thus evident that the invariance of the vanishing CBC under the $T$ transformation is ensured by the appearance of the affine contribution $\Xi$ in~\eqref{Atilde}. Explicitly, this affine term, in accordance with \eqref{Xi}, is given by
\begin{eqnarray}
    \Xi&=&i\langle\psi^R_-|\eta(\Gamma'-\Gamma)|\psi^R_-\rangle = i\langle\psi^R_-|\eta \left(T\partial_\lambda T^{-1}\right)|\psi^R_-\rangle \nonumber \\
    &=&\frac{i}{2}\langle\psi^R_-|\eta \begin{pmatrix}
        \tanh\xi & 0 \\
        0 & \tanh\xi
    \end{pmatrix}|\psi^R_-\rangle \nonumber \\
    &=&\frac{i}{2}\tanh\xi\langle\psi^R_-|\eta|\psi^R_-\rangle=\frac{i}{2}\tanh\xi.
\end{eqnarray}
As a result, this affine term exactly compensates the imaginary component that appears in the conventional left–right Berry connection in~\eqref{A'lr}, thus ensuring that the norm of the wave function is automatically preserved when the state is adiabatically transported in the system parameter space.

\renewcommand{\theequation}{D\arabic{equation}}
\setcounter{equation}{0}
\section{Hermitized Hamiltonians and their eigenstates}\label{AC}
The fact that $\mathbb A=0$ can be alternatively understood from the perspective of Hermitized Hamiltonians and their eigenstates. Since the metric, and hence the Hermitizing map, is time independent, which is always the case in the unbroken pseudo-Hermitian phase of the Hamiltonian, the associated Hermitian Hamiltonian can be written as $H^{H} = S H^{\rm pH} S^{-1}=-l\sigma_x$, with the corresponding eigenvectors given by $|\phi^H_i\rangle = S|\psi^R_i\rangle$. Explicitly, one finds: 
\begin{eqnarray}\label{phiH}
|\phi^H_-\rangle = \frac{1}{\sqrt{2}}
\begin{pmatrix}
1 \\ 1
\end{pmatrix},
\qquad
|\phi^H_+\rangle =
\begin{pmatrix}
-1 \\ 1
\end{pmatrix}.
\end{eqnarray}
Similarly, one obtains the Hermitizing map for the rescaled eigenvectors as follows
\begin{eqnarray}\label{S'}
    S'=\sqrt{\cosh\xi} S.
\end{eqnarray}
That is, the new Hermitizing map is just the `rescaled' map $S$. Consequently, the associated eigenvectors $|\phi'^H_i\rangle$ are equivalent to the eigenvectors $|\phi^H_i\rangle$ in \eqref{phiH}. 
As a result, the transformed CBC $\mathbb A'$ is identically zero.

\bibliography{references}
\end{document}